**Title**

Thirty years of molecular dynamics simulations on posttranslational modifications of proteins

Austin T. Weigle[1], Jiangyan Feng[2], and Diwakar Shukla[2,3,4,5]*

[1]Department of Chemistry, [2]Department of Chemical and Biomolecular Engineering, [3]Center for Biophysics and Quantitative Biology, [4]Department of Bioengineering, [5]Department of Plant Biology, University of Illinois at Urbana-Champaign, Urbana, Illinois 61801, United States

E-mail: diwakar@illinois.edu

**Abstract**

Posttranslational modifications (PTMs) are an integral component to how cells respond to perturbation. While experimental advances have enabled improved PTM identification capabilities, the same throughput for characterizing how structural changes caused by PTMs equate to altered physiological function has not been maintained. In this Perspective, we cover the history of computational modeling and molecular dynamics simulations which have characterized the structural implications of PTMs. We distinguish results from different molecular dynamics studies based upon the timescales simulated and analysis approaches used for PTM characterization. Lastly, we offer insights into how opportunities for modern research efforts on *in silico* PTM characterization may proceed given current state-of-the-art computing capabilities and methodological advancements.

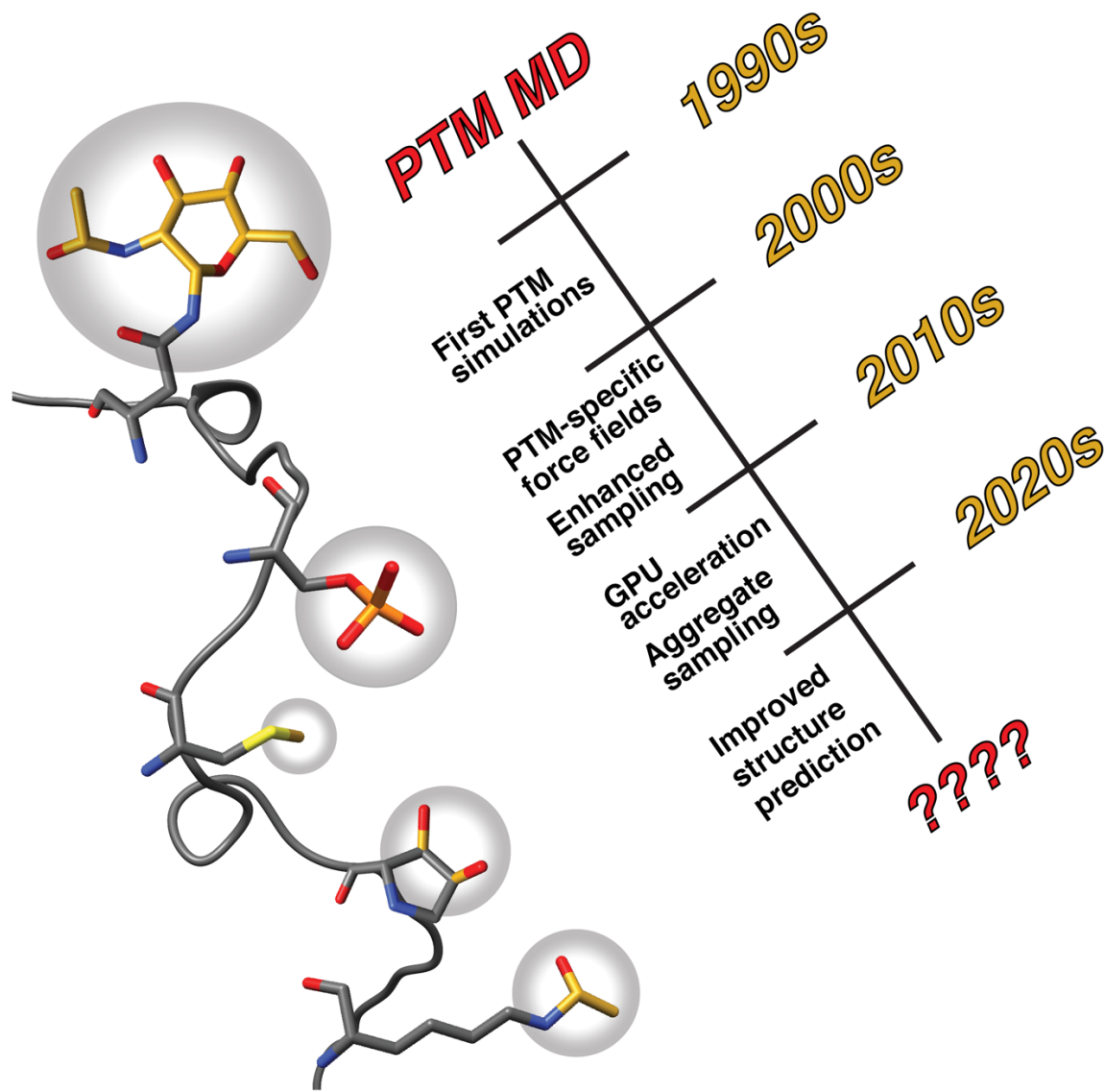

**Figure 1.** As of 2022, computational studies of the structural effects induced by posttranslational modifications have been performed for nearly three decades. Here, we review what is feasible when simulating posttranslational modifications given previous and current state-of-the-art modeling and analysis techniques.

## 1. Introduction

The basis of cellular function is rooted in atomistic movement. Irrespective of the molecular form, atomistic interactions resulting from these movements signal the need for necessary biophysical actions. DNA replication, RNA transcription, protein translation, and related signal transduction are all fundamental molecular processes which are dynamically recruited to satiate cellular need. Accordingly, an organism's genomic architecture equips its cells with the tools necessary to maintain homeostasis. The genome encodes genetic and proteinogenic responses to cellular perturbation. However, genomic architecture fixture occurs on evolutionary timescales. Thus any *a priori* information available to cellular machinery via the genome may not always possess the precision nor efficiency for any given stimulus occurring along physiological and developmental timescales. True, DNA and protein synthesis are essential provisions for maintaining homeostasis and cellular feedback mechanisms, but additional forms of spatiotemporal regulation are required to guarantee that the cell can acclimate to emergent complexity.[1] In turn, cells developed a versatile biochemical arsenal through posttranslational modifications (PTMs).

PTM refers to either reversible or irreversible chemical changes in proteins after translation, which often occur on specific amino acid side chains.[2] PTM addition or removal is executed by means of a covalent modification by bonding or hydrolytic cleavage, typically occurring on side chains which may behave as nucleophiles (Cys, Ser, Thr, Tyr, Lys, His, Arg, Asp, Glu).[3] Hydroxylation of Pro and Asn act as exceptions.[3] Approximating almost 700 unique entries in the UniProt database to date,[4] the most common additive PTMs are phosphorylation, acylation (includes ubiquitination), alkylation, glycosylation, and oxidation; less common ones include hydroxylation/carboxylation, the addition of peptide moieties to preexisting protein residues, and sulfur-sulfur transfers, among others.[1,5,6] Proteolysis, deamidation, and eliminylation are examples of nonadditive PTMs.[7] Some PTMs for commonly modified residues, which can be computationally modeled, are shown in Figure 2. Each PTM serves a role in altering the target protein in some way, perhaps by introducing stability or redirecting localization and subsequent interactions post-installment.[8] In effect, target protein structure and function are inherently changed upon PTM introduction, as the PTM fundamentally changes the physicochemical properties of its host amino acid.[3,5]

Thousands of proteins exist within individual proteomes, comprising a set of machinery capable of "reading", "writing", and "erasing" PTMs.[10–12] Reader proteins recognize PTMs installed by writers (kinases, ubiquitin ligases, acetyltransferases), through molecular interactions for the preparation of an adequate biological response to cellular conditions.[12] For reversible PTMs, eraser proteins (phosphatases, deubiquitinases, deacetylases) terminate molecular signal propagation by removing PTMs.[12] Furthermore, target proteins may contain multiple PTM sites, resulting in combinatoric PTM possibilities which may redefine expected cellular outcomes that would otherwise occur from just a single PTM instance.[1] Although PTMs may instantiate a sense of proteomic crosstalk, they are also capable of intra-protein crosstalk, communicating allosteric information about what kind of conformation the target protein should adopt in preparation for its next role in cellular signaling and stress response.[13] Interrogating how PTMs communicate such intra- and inter-protein changes thus becomes central to understanding how protein function is regulated throughout any physiological state of the cell.

As experimental advances in mass spectrometry and chemical biology have facilitated PTM

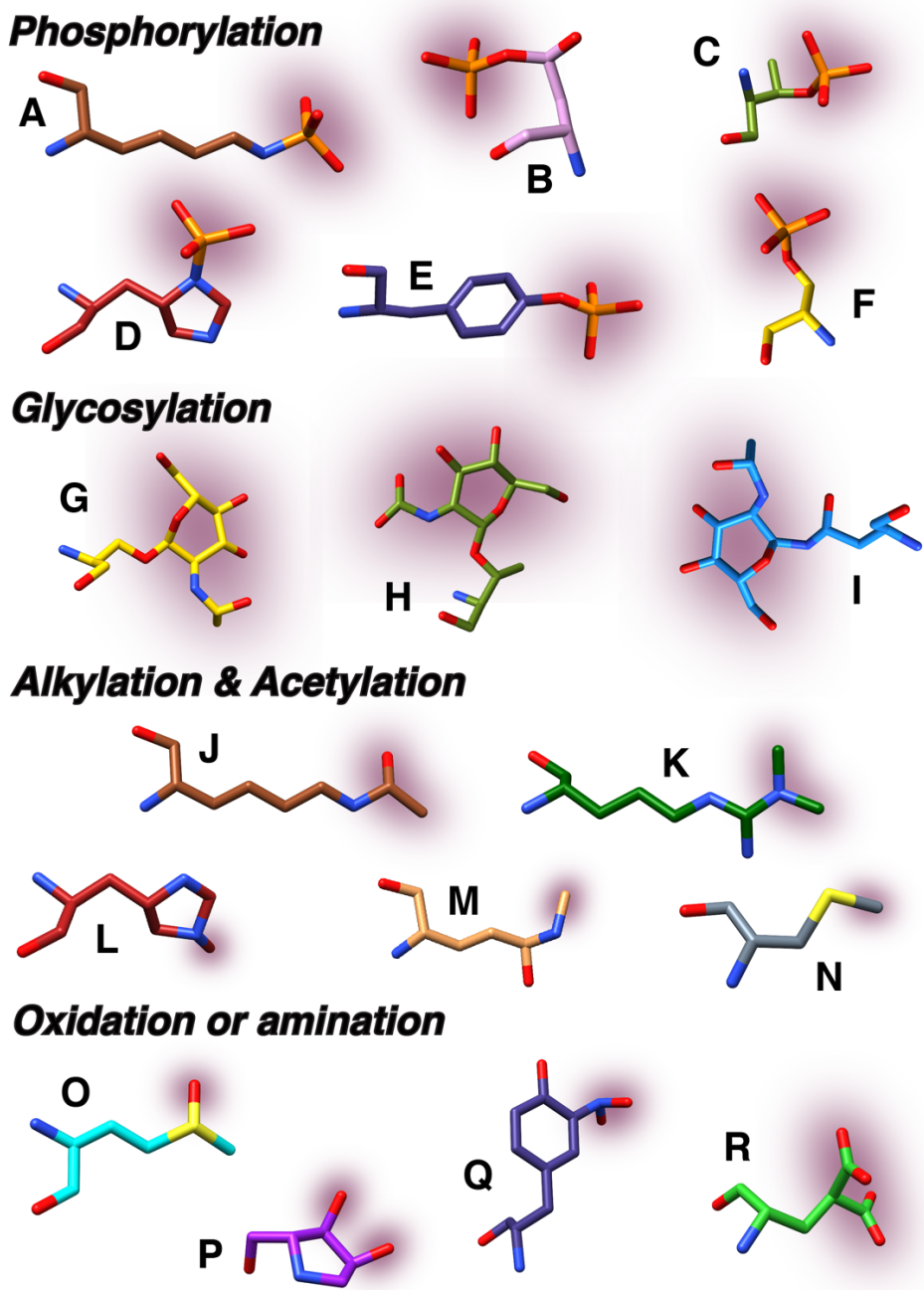

**Figure 2. Display of modeling capabilities for common PTMs on commonly modified residues.** Modifications are shown for residues which can be readily modeled using the Vienna-PTM server.[9] PTMs are highlighted with a black raspberry glow. Structures resulting from the server are only parameterized with GROMOS force fields. Phosphorylation is shown for (A) lysine, (B) aspartate, (C) threonine, (D) histidine, (E) tyrosine, and (F) serine. Glycosylation is shown for (G) serine, (H) threonine, and (I) asparagine. Acetylation is shown for (J) lysine. Alkylation is shown for (K) arginine, (L) histidine, (M) glutamine, and (N) cysteine. Oxidation is shown for (O) methionine, (P) proline, and (R) glutamate. Amination is shown for (Q) tyrosine.

discovery and methodologies for characterization,[12,14,15] the opportunity to systematically describe PTM functional outcomes gains clarity. Still, barriers exist to an exhaustive experimental interrogation of how PTMs influence protein function. Systematic studies using mass spectrometry often consider identification for few modification types of a single PTM rather than considering the existence or functional consequences of multiple PTMs.[2] Exploration of multisite PTM characterization poses similar challenges as seen for exploring higher-order mutations in variant effect prediction with respect to throughput.[16] Although cellular assays can be performed to examine PTM effects on protein expression, function, and downstream effects, dissecting how these modifications relate to individual protein structure and conformational changes requires more effort. Throughput for spectroscopy and structure determination experiments is limited by large scale efforts of *in vitro* heterologous expression for both soluble and membrane proteins, as well as specialized training for probe/label design or preventing protein precipitation. However, *in silico* approaches to studying PTMs offer an alternative for challenging experimental characterization efforts.

In this Perspective, we highlight the foundations laid by PTM *in silico* research spanning the last 30 years to offer insights towards future PTM characterization using computation. Recent advances in computing power and simulation methodology have rendered *in silico* molecular modeling and molecular dynamics (MD) simulations an important, and increasingly accessible, tool to understand biological macromolecules.[17–21] Here we compare what MD simulations of PTMs have achieved in the past using less computational power and fewer methodological advances with what is now capable with more modern resources and simulation techniques. Our goal is to illuminate the eventual quality of PTM functional understanding so that it may rival the increasing quantity of discovered PTMs.

## 2. Modeling posttranslational modifications for molecular dynamics simulations

*2.1. Introducing molecular dynamics simulations*

MD simulations require two inputs: initial positions extracted from an experimentally determined (or computationally predicted) structure and a force field describing the interactions between atoms. A typical force field consists of the following terms: covalent bonds, bond angles, dihedral angles, electrostatic interactions, and other non-bonded interactions. According to Newton's second law of motion, the acceleration of each atom can be computed. Using numerical integration techniques, the positions and velocities of every atom in the system can be updated as a function of time. To ensure numerical stability, the time steps in an MD simulation must be short – typically only a few femtoseconds – so that each time step is shorter than the timescale of the fastest chemical bond vibration. The output of an MD simulation is a trajectory file containing the position of each atom at each time step, describing the motion of the system at the atomic level during the simulated time period.[22] Preliminary stages of simulation involving heating and restraint relaxation are applied first, ensuring thermostability of modeled molecules and their starting conformations. After undergoing short simulations where restraints have been removed and the model system has converged to a stable state, data can then be reliably recorded from trajectories known as "production runs" with minimal fear of simulation artifacts.[23]

Before committing to any computational study design, the greatest bottleneck to conducting MD studies is the availability of computing resources. This caveat is especially important, as the simulation of a given biomolecular process must occur over a specified timescale to accurately sample the conformational ensemble for the dynamics of interest.[18,24,25] Scientific computing costs have been alleviated as more powerful hardware has been produced, but simulation throughput has historically not been as efficient nor parallelizable as can be seen today. Within the first decade of the 21st Century, the acceleration of MD simulations via porting to graphical processing units (GPUs) was retarded by a few lurking variables – namely scaling across multiple GPUs, changing how memory was accessed on GPUs versus central processing units (CPUs), and interfacing with compilers which would allow for MD software to run on GPU source code.[26] By 2010, these challenges were eventually overcome by software developers, where MD simulation work performed on GPUs became routine and continuous efforts were made to increase simulation efficiency.[27–30] Naturally, the outcomes of GPU porting achievements were amplified by the integration of MD workflows onto large distributed computing as well as supercomputing platforms (e.g., private clusters, Folding@Home, D.E. Shaw's Anton, the National Science Foundation's Frontera, or National Center for Supercomputing Resources like BlueWaters and XSEDE initiatives).[31–36]

MD methodologies can be applied for studying the evolution of specific biomolecular processes, evaluating how these processes may be perturbed following a controlled change, or simply demonstrating conformational flexibility and stability along some molecular structure.[19] Technical advances for MD simulation can be divided based on whether they address phase space sampling or interpretation of resulting simulation data. While we focus on MD study design with respect to PTMs later in this Perspective, the different types of MD simulation techniques, their application and development, have been extensively reviewed elsewhere.[37] On the other hand, the large deluge of results and calculations derived from MD simulations has prompted the use of machine learning for improved data representation and analysis.[38–40]

Regardless of the method used for sampling, measures for uncertainty quantification and sampling quality have been communicated as best practices when designing and conducting MD simulation studies.[41] A variety of simulation techniques used to calculate binding free energies, thermostability, or kinetic rates of transition often come in exact agreement, or occur within error, of experimental observations.[42–47] Minimally, MD simulation can offer qualitative agreement with experimental results. Simulation accuracy is extremely sensitive to the initially modeled conditions, but can be strengthened by repeated and systematic sampling of the phase space for a given process. Outside of the modeled structure and selected parameterization set quality, the predictive power of MD simulations and the minimization of error lies in the attempt for approximating statistical convergence of ensemble-averaged properties.[48] Today, this is more easily achieved by running longer and greater trajectories thanks to better computing and ensemble simulation-based approaches.[48]

*2.2. In silico installation of PTMs for preparing simulation studies*

Proteins are highly dynamical systems with functions governed essentially by their dynamical behaviors.[49] MD simulations thus become a powerful tool for studying PTMs. Notably, MD simulations can capture the dynamic behavior of biological systems at atomistic resolution in a label-free manner. They can therefore offer a unique structure-dynamics perspective in deciphering the roles of PTMs on proteins without any form of perturbation introduced by probes.[50,51] Furthermore, the simulation conditions can be carefully controlled by the simulation practitioner. So by comparing simulations performed under different PTM conditions, one can identify the effects of a wide variety of PTMs, which is usually experimentally unfeasible.[52] Accurate PTM modeling is still reliant on appropriate information related to the PTM-bearing amino acid site and tools to introduce and parameterize these changes to the protein structure.

Databases and informatic methods cataloguing general PTM information,[53–65] as well as resources depicting experimentally-validated or computationally-predicted sites specific for phosphorylation,[66–87] glycosylation,[79,88–97] and acylation,[79,98–118] among other modification forms,[79,119–131] exist. These resources can be used to identify PTM sites on target proteins for improved molecular modeling. Traditionally, if a PTM was not already present on a resolved crystal structure, covalent modifications were made to input coordinate and topology files to reflect the correct modification using modeling software or through manual text editing of the input files. Such hands-on changes can be introduced using either molecular visualization software[132–137] or system preparation modules found in popular MD simulation engines (e.g. AMBER, CHARMM, GROMACS, NAMD).[28,29,138,139] Eventually, standalone- and web-tools were developed to streamline PTM installation onto protein structures.

As a historiographic example, glycosylation efforts starting from the late 1990s and continuing into the mid 2010s were focused on modeling carbohydrate structures, although none of these tools could directly merge the resulting sugars onto protein coordinates and generate glycoprotein topologies necessary for simulation.[140–144] That is, these software offer limited accessibility to PTM modeling by making simulation topology generation exceedingly nontrivial for nonexperts. In 2005, GLYCAM emerged as a tool developed specifically for modeling glycoproteins and solution carbohydrates as AMBER and CHARMM simulation inputs.[145] CHARMM developers then introduced glycan reading and modeling functionalities into the CHARMM-GUI webserver for glycoprotein structure preparation.[146–148] Similarly, GROMACS developers made *doGlycans* for modeling GROMACS-compatible glycoprotein structures.[149] Although each of these tools enabled glycoprotein structure preparation, preference of one MD engine over another confound progress on PTM modeling. Features available in some tools for one engine – i.e., sugar library chemical and conformational diversity – may not be available for another, requiring nontrivial file conversion and topology preparation strategies using additional or third-party tools. Indeed, juxtaposed singularity between different software and computational tools is an inherent obstacle in molecular modelling!

By 2017, several MD engine camps had developed their own respective tools for glycoprotein structure generation, inspiring force field-independent efforts to make carbohydrate structures available and amenable for MD input file preparation.[150–152] Outside of common PTMs like phosphorylation and ubiquitination, additional tools were developed to facilitate modeling. The PyMOL plugin PyTM can model phosphorylation, acetylation, carbamylation, citrullination, nitration, methylation, hydroxylation, aldehyde adduct formation, as well as any combination of the above.[153] The Vienna-PTM server offers residue templates for over 256 (non)enzymatic modifications, although modified output structures are designed to be used within a GROMACS framework.[9] Despite these advances, CHARMM-GUI maintains popularity because of its ease of access; its phosphorylating, lipid anchoring (acylate), glycosylating, and peptide stapling capabilities; and its file output schemes that maximize compatibility with diverse MD engines (Figure 3).[154–156] Regardless of the methods used for PTM installation, modified amino side chain orientations may be further optimized using rotamer libraries or through equilibration simulations.[157]

Figure 3. Modified screenshot of CHARMM-GUI PDB Reader function highlighting specific PTM options for input protein structure manipulation.

With protein structure files covalently modified, PTM parameterization is required to ensure that simulations accurately depict biophysical behaviors as expected from experiment. Initially, PTM models did not necessitate their own specific force fields. MD practitioners could then argue that accuracy of PTM-containing simulations was not significantly compromised, as atom parameters could be assumed by analogy from preexisting libraries. Such practice was acceptable at the turn of the 21$^{st}$ Century, before more PTM-specific force fields were developed. For common PTMs, phosphorylation moieties could borrow nucleotide parameters to describe their phosphate groups; likewise, glycosylation, ubiquitination, and acylation additions shared atoms with protein and small molecule force fields. Missing parameters were otherwise developed as needed by specific research groups without immediately being publicly available. Specific force field developments for particular PTMs, such as phosphorylation and glycosylation, were later expanded into addons and were incorporated into libraries commonly used for each of the popular MD engines.[158–168] It is worth noting that the majority of discussed modeling and parameterization efforts were first dedicated to characterize biologically common PTMs, suggesting that a need for improved modeling of less common PTMs still remains.

**3. Designing molecular dynamics approaches for the study of posttranslational modifications**

Perhaps the beauty of molecular simulation lies in its unfettered power to describe molecular systems in almost unlimited detail. Extent of available computational resources and methodologies will always be a bottleneck to data procurement, but the onus of appropriate research design and meaningful analyses falls to the MD practitioner.

Generally, trajectory analyses assume either a time-dependent or time-independent form. Time-dependent data collection regimes incorporate analyses that project some feature against the order parameter of time, following observations exactly as they transpired along individual trajectories. In contrast, time-independent data collection regimes instead employ analyses that project data against order parameters as a timeless ensemble, where data points occupy positions within a dimensional phase space with respect to the selected order parameters.[169–172] Removing reliance on time allows for data to be viewed from a perspective of population density as opposed to mere frequency of occurrence within a trajectory. Such a style of data projection can be seen in free energy landscapes (FELs), where aggregate data points occupying highly frequent regions in the projected dimensional space can be identified as either metastable or transitionary macrostates.

Each analysis and data collection approach has its strengths and weaknesses. Time-dependent data collection and analyses are typically employed to focus on specific interactions given a starting structure and with or without a target structure in mind. These types of results follow "single" or "few" trajectory simulation schemes where coordinate and velocity information are retained to maintain the existence of a long trajectory. As such, time-dependent data representations offer a simplified perspective reinforced by multiple replicates of production runs observing similar molecular events. Because data output by individual time-dependent trajectories is not aggregated, enhanced sampling methods are often used instead of just classical MD simulations to capture molecular transitions occurring along longer timescales. Approaches like accelerated, targeted, or steered MD are examples of such techniques.[173–175] Although studies following a time-dependent data collection regime can utilize time-independent analyses, these studies tend to lack sufficient sampling to offer comprehensive conclusions beyond the direct research question in focus.

On the other hand, time-independent data collection and analyses are employed to observe a desired transitionary path from one macrostate to another along order parameters which are not based in time. Data representation can then naturally take the form of FELs. Studies designed using time-dependent approaches are still capable of capturing a FEL but may not be able to do so for longer timescale processes. Time-independent approaches achieve this goal by combining the results from an aggregation of "brute force" classical MD simulations or by biasing potentials along selected order parameter(s). Such a biasing method includes metadynamics, where the choice of order parameters.[176] Results from brute force simulation of unbiased parallel trajectories need to be "stitched" together using kinetic frameworks like Markov state models (MSMs) or milestoning,[177–180] while path search and optimization protocols can otherwise be biased along potentials using approaches like umbrella sampling, replica exchange, or the string method.[181–183] Although nontrivial to implement, results from unbiased and biased simulations can be augmented into a single multiensemble kinetic framework for time-independent data representation.[47] Depending on the extent of collective variable exploration, time-independent data representation can demand extremely more resources than research designs relying on a time-dependent data collection and analysis approach.

Herein, we describe the evolution of MD studies on PTMs and their effect on protein structure-function. We begin by discussing methodologies and results from studies observing how PTMs influence protein structure and function along shorter timescales. In these studies, the dominant styles of molecular simulation include time-dependent data collection and analysis regimes, as well as path optimization methods like replica exchange MD (REMD) given available experimental structures. Subsequently, we transition to studies employing FELs for detailing how PTMs impact the range of protein function. Thanks to increased computing capabilities and methodological advances, PTM MD studies wielding a FEL-based approach have recently become more widespread within the literature.

## 4. Foundational insights on posttranslational modifications revealed by studies using trajectory-specific approaches

During the mid-1990s, MD practitioners were first challenged with the task of modeling and simulating covalently modified proteins. At first, it was questioned whether results from PTM simulations could even concur with experiments. The initial quest to address how PTM dynamics regulated protein function was confounded by other lurking variables, too. As described in Section 2, few existing tools to easily modify protein structures were available; therefore, initial works were constrained to studying proteins whose resolved structures already contained PTMs. Additionally, GPU-based acceleration of MD simulations was not implemented until the late 2000s, meaning that approaches to modeling PTMs needed to be carefully designed so that they could efficiently offer unique insights with minimal data and computing time requirements. As such, many of these initial simulation studies between 1990 and 2010 were performed along picosecond to nanosecond timescales, focusing on specific regions of protein structures or interactions with covalently modified peptides. Short timescale MD studies were capable of modeling intramolecular protein sidechain contacts and whether target protein or peptide structures were altered by PTM installation. Identifying how hydrogen bonding networks could be altered after PTM introduction was a worthwhile contribution for the time, as NMR experiments could not always distinguish hydrogen bonding interactions involving PTMs which lacked hydrogens. Despite how early computing capabilities limited MD to the study of short timescale transitions like changes in hydrogen bonding, any atomistic details offered from experiments alone were still incomplete without the help of molecular modeling.

*4.1. Characterization of active and binding site rearrangement through PTM interactions*

What was first of interest included the structural consequences of PTMs on sidechain orientation and peptide secondary structure within binding pockets and recognition interfaces. In 1995, the basis of aldolase inhibition by a synthetic peptide matching the first 15 residues of human erythrocyte peptide band 3 (B3P) was characterized using MD simulation of NMR-resolved aldolase structures.[184] The structure of the peptide in complex with aldolase was determined via NMR, and a phosphorylated version of the peptide was then docked into the aldolase binding pocket and simulated for 5 ps. An unphosphorylated hydroxyl moiety of the peptide's tyrosine group was able to experience favorable interactions within the docking site, but pTyr resulted in unfavorable interactions and inhibited certain rotational ranges of motion because of electrostatic repulsion. These findings offered an explanation as to how B3P phosphorylation inhibited glycolytic function of aldolase off phosphate electrostatics alone.[184] Using computing capabilities available before the year 2000, the binding of unmodified, $P_i$-bound and phosphopeptide-bound phospholipase SH2 domain MD simulations were modeled in a separate study.[185] To give perspective on the available computing power, SH2 was simulated for 250 ps at a rate of 3 ps/day in this study.[185] These phospholipase simulations demonstrated how PTMs recruit mostly the same residues as their solution-based counterparts (e.g., $P_i$ versus pTyr), as $P_i$-bound and pTyr simulations highlighted the importance of a conserved arginine triad in stabilizing phosphate group positioning within the SH2 binding pocket (Figure 4).[185] Here, short simulations could suggest an evolutionarily conserved role within binding pockets for select residues when handling PTMs by maintaining the hydrogen bonding network of the active site.

In other cases, there exists a possibility for active site rearrangement following PTMs, which need not necessarily be a result of direct residue-residue interactions with the covalently modified PTM

**18**

site. Comparison between catalytic domains seen in the phosphorylated structures of inactive (closed) Src tyrosine kinase and active (open) Lck kinase suggested that phosphorylation-induced

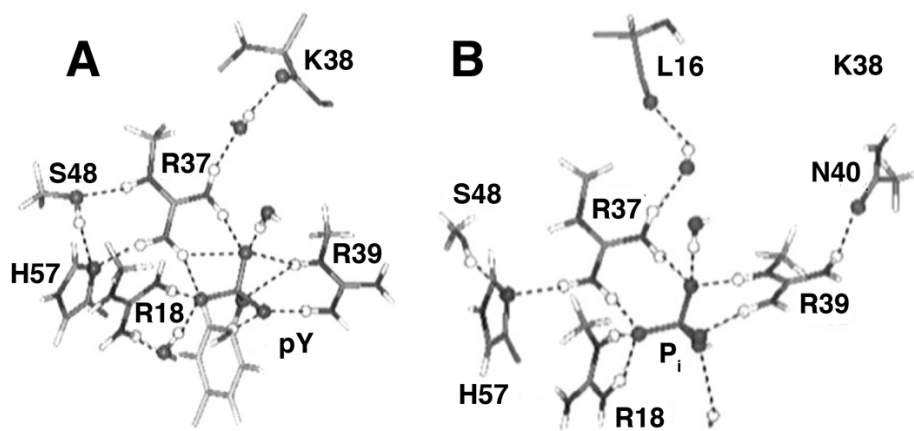

**Figure 4. SH2 domain phosphate binding pocket hydrogen bonding network.** (A) Hydrogen-bonding network shown for phosphopeptide binding to the phosphate binding site. (B) Hydrogen-bonding network for inorganic phosphate to the phosphate binding site. Both binding simulations depict central role of conserved arginine triad in stabilizing phospho-moieties. Adapted with permission from M.-H. Feng, M. Philippopoulos, A. D. MacKerell and C. Lim, *J. Am. Chem. Soc.*, 1996, **118**, 11265-11277.[185] Copyright 2022 American Chemical Society.

conformational change along a polypeptide linker was required to reflect kinase active state (Figure 5).[186] To study this dynamic hinge motion hypothesized for maintaining active versus inactive states upon phosphorylation, a targeted MD approach was used where increasing amounts of harmonic restraints were applied across increments of 60 ps to the Src kinase hinge region Cα atoms.[186] From this, targeted MD converted pTyr416 Src kinase from the resolved closed state to a target pseudoactive open state based off the Lck crystal structure.[186] Targeted MD could not exactly mimic an Lck-like extended conformation until bound ATP was replaced with ADP, which then forced immediate conformational change within 800 ps in support of experimentally proposed intramolecular autophosphorylation of Tyr416. Thus, in this 2004 study, Src kinase could only achieve the active conformation when the active site reorganized around pTyr416 and ADP

**19**

interactions.[186] Similar perturbations to cofactor coordination in cyclin-dependent kinase 2 (CDK2) active site residues were seen during 60 ps classical MD simulations of inhibitory phosphorylation along the G-loop.[187] Still, the position for phosphorylation is critical for determining the effect on active site rearrangement. For CDK2, it was reported from classical simulations ranging between 3-10 nanoseconds that pThr160 activation occurred by constraining the active site T-loop for phosphotransfer; meanwhile pThr14 and pTyr15 introduced enough flexibility in the G-loop to widen the active site, destabilize cofactor placement, and inhibit phosphotransfer.[188,189] Thus, PTM installation can rearrange loop conformations to alter protein function by generating new cavities for accommodating protein-specific activation functions.[190]

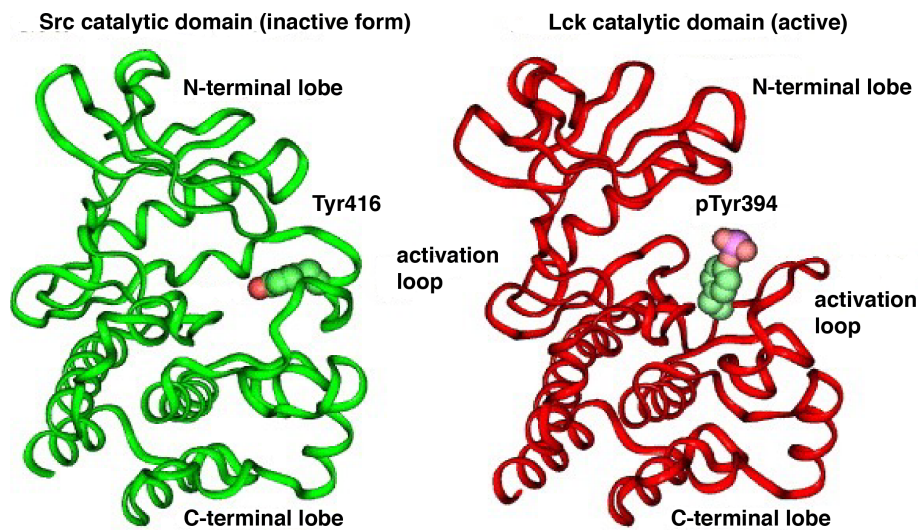

**Figure 5. Comparison of Src kinase inactive structure (left) versus Lck active structure (right).** A common strategy for exploring PTM-induced conformational change at reduced computational cost was to use covalently modified starting structures in different conformations. This case study used the Lck kinase phosphorylated active structure to create a homology model for the active Src kinase structure. Adapted with permission from J. Mendieta and G. Gago, *J. Mol. Graph. Model.*, 2004, **23**, 189-198.[186] Copyright 2022 Elsevier.

Active site reorganization can also occur through solvation effects. Phosphorylation of human eukaryotic translation initiation factor 4E, or eIF4E, found that pSer209 narrowed the dinucleotide cap binding pocket after shortening nearby backbone Cα-Cα distances by about 10 Å throughout an 800 ps simulation performed in 2003.[191] However, this tightening of the dinucleotide binding pocket could be accredited to dianionic phosphate recruiting a large solvation shell as a consequence of the pSer209 phosphate group's electronegative charge. This localized electronegativity thus created a force that pushed active site residues away and further enclosed the dinucleotide binding site.[191] Either by electropositive amino acid composition or hydration, positive charge is needed in order for phosphopeptide binding or phosphorylation to be supported.[192] In the case of phosphorylation, constrained conformational change is likely the means by which proteins accommodate the introduction of negative charge.

*4.2. Influence of PTMs on protein structure, thermostability, and allosteric communication*

Outside of active site studies, other short peptide simulations would proceed to computationally characterize the effects of phosphorylation by highlighting the importance of electronegativity introduced by phosphorylation. Phosphorylation was seen to induce conformational changes in a 10-residue region of tyrosine hydroxylase where most metastable states depicted peptide compression after 5 ps of simulation, unlike what was seen in the unphosphorylated structure.[193] Salt bridges seen in the unphosphorylated tyrosine hydroxylase peptide were broken, leaving many charged residues solvent-exposed as the pSer moiety hydrogen bonded with terminal residues. New interactions and conformations caused by Ser phosphorylation at one tyrosine hydroxylase site were argued to be the cause for cooperative increases seen in experimental phosphorylation at additional sites along the peptide region.[193] Peptide compression in response to phosphorylation has been reported in additional studies spanning from 2002 to 2012, too, as well as the breaking of native salt bridges in order to generate hydrogen bonds for mitigating the negative charge of phosphate groups.[194,195] Other forms of conformational change for Ser phosphorylation have been reported, such as stabilized alpha-helical secondary structure formation based on phosphate-backbone electrostatic interactions,[196] as well as preferred dihedral orientations introduced at the phosphorylation site.[197] Thr-specific phosphorylation was also seen to increase helicity in 100 ns classical simulations of p53 and p73 proteins, although decreases in helical content have been seen for other proteins.[198,199] Even as recent as 2019 has the interplay of individual PTM forms been found to stabilize N-terminal helices through phosphate charge neutralization and backbone-sidechain hydrogen bonding via advanced REMD simulations.[200] From the large body of phosphorylation modeling and simulation research, modification along specific amino acids and in different local environments will alter the probabilities of distinct resulting conformations. Work from the Jacobson group using umbrella sampling and quantum mechanical studies found in 2007 that hydrogen bond strength with phosphate groups noticeably changes in response to the geometry assumed by different phosphorylated residues based on their amino acid identity and their local environments.[201] Similarly, REMD simulations on glycosylated proteins also showed in 2010 that protein secondary structure can be altered in a site-specific manner based on shifted backbone dihedral preferences after glycosylation.[202]

Intrinsic disorder and repositioning of flexible loops serve as strategies for proteins to rapidly change their local environments in response to dynamic processes. PTM installation along disordered regions can instead exacerbate cellular perturbation response strategies of proteins by altering thermostability.[203] In 2008, the Levy group designed an elegant study by systematically

**22**

modeling glycan groups onto an SH3 domain protein to distinguish the entropic versus enthalpic effects of glycosylation on protein function (Figure 6A).[204,205] Coarse-grained modeling of the glycosylated SH3 structures enabled this systematic study design, where 63 different glycosylated SH3 domains could be efficiently simulated at different temperatures and in replicates. By varying temperature for the different simulations, Arrhenius-style free energy projections showed how increased extent of glycosylation increased protein thermal stability, where both enthalpy and entropy were reduced by the addition of glycans (Figure 6B-D). The addition of glycans to SH3 was found to rigidify the protein structure, thereby lowering the enthalpy as the glycan moieties restrict protein dynamics. Given the breadth of collected simulation data for the time, Shental-Bechor and Levy's study could offer direct insights towards SH3 stabilization that could not be offered by experiments alone.[204,205]

Even if PTMs may not principally operate through electrostatic perturbations, the entropic and enthalpic contributions offered by a specific covalent modification contribute to how the effects of a PTM manifest in a site- and modification-specific manner. Modeling and simulation by the Levy group have accurately predicted glycosylation-induced destabilization of WW-domain protein in regions where large numbers of contacts in native topology models preexist, as glycan installation would disrupt otherwise stabilizing interactions.[206] However, the opposite is not necessarily true when comparing simulation to experiment. The same 2010 study found that just because a glycosylation site along the WW-domain scaffold appears as if it can sterically accommodate a glycan does not universally imply that such a modification will result in stabilization.[206] Entropic stabilization could instead arise through PTM-based dehydration, as WW-domain protein PEGylation simulations suggest that solvent removal from the protein surface

**23**

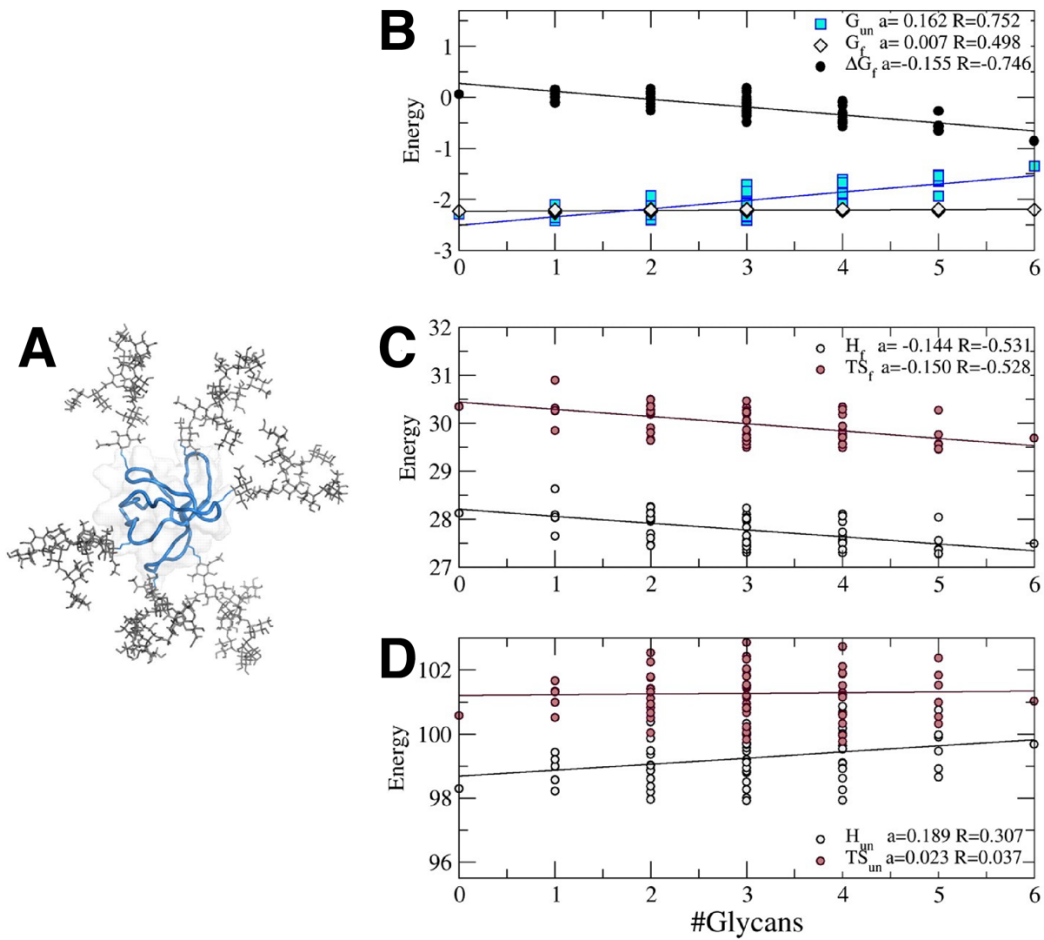

**Figure 6. Glycosylation effect on thermodynamic stability of SH3 domain protein. (A)** SH3 domain represented as a blue ribbon, with glycosylation moieties represented as grey sticks. This modified protein was simulated using a coarse-grained native topology model rather than the all-atom version presented here. **(B)** Folding free energy ($\Delta G_f$) of the SH3 domain protein. **(C)** SH3 domain protein folded state thermodynamics, broken down by enthalpic and entropic contributions. **(D)** SH3 domain protein unfolded state thermodynamics, broken down by enthalpic and entropic contributions. Panels **B-D** represent free energy as a function of the length of the glycan chains added. Adapted with permission from D. Shental-Bechor and Y. Levy, *Proc. Natl. Acad. Sci. U.S.A.*, 2008, **105**, 8256-8261.[204] Copyright 2022 National Academy of Sciences.

by PEGylation improves solubility and folded state preference.[207] Thus, it can be expected that entropy-based PTM effects are tied to the hydrophobicity and stericity inherent to that modification. In 2016, the Jacobson group found that the entropic effects of ubiquitination on a human kinase could not be replicated when performing acetylation at the same site, as an active-like conformation caused by ubiquitination was not observed across 10 ns simulations of an acetylated structure.[208] The difference in effect between acetylation versus ubiquitination could be attributed to the difference in size and bulk between the two types of covalent modification, a finding which likely applies to other forms of PTMs as well. For instance, 2020 modeling and MD simulation of diverse PTMs onto the central coiled-coil connector of fibrinogen varied from having zero to severe structural effects.[209]

The impacts of PTMs on protein conformation extend beyond local environments and into long-range contacts and allosteric communications. As the Levy group helped identify how loss in natively stabilizing interactions upon glycosylation underscored structural destabilization in 2015, they further showed how interactions lost from glycosylation also include long-range contacts in the case of simulations of glycosylated MM1 protein.[210] Additionally, long-range communication between covalently-modified residues is also site-specific, as only modifications on certain regions of a protein may induce conformational change in regions distinct from the site of modification.[211] For example, terminal phosphorylation only increased cross-correlated motion between the SH2 and SH3 peptide binding domains in c-Src kinase, but not for other parts of the protein during simulations from 2001.[212] In the case of multiple PTM sites, only acetylation on a specific H4 histone tail lysine residue uniquely altered the simulated conformational ensemble in 100 ns REMD simulations, although acetylation at any lysine along the tail increased long-range backbone contacts.[213] Synchronous residue-residue contacts can also been influenced by PTM crosstalk. Different REMD system setups modeling combinatoric acetylation of High Mobility Group Box (HMGB) protein lysines found that contact networks in HMBG:DNA complexes differed despite each combination being minimally comprised of the same two to four lysine site acetylations.[214]

*4.3. Summary of PTM simulations using "single" or "few" trajectory approaches*

Despite using short trajectories analyzed in time-dependent manners, most of the works discussed above illustrated key findings with respect to how PTMs impact protein dynamics and function. Binding of covalently modified peptide substrates indicated direct interactions between protein binding interfaces, where side chain reorientation accommodated new electrostatic or steric forces brought by the PTM. When installed onto part of the target protein structure, PTMs were seen to alter secondary structure of disordered regions by affecting backbone hydrogen bonding preferences and again calling for side chain reorientation to offset any destabilizing interactions with stabilizing ones. Conformational preference thus reciprocates the nature by which a protein structurally responds to a particular PTM. Subversive changes in local side chain environments in response to changed hydrophobicity or electrostatic potential nearby the PTM site are perhaps the driving forces which help guide a modified protein towards a metastable state. Systematic modeling of PTMs indicative of yeast stress responses have shown that covalently modified structures simulated for 20 ns classical MD simulations do demonstrate conformational change, but that the extent of conformational change largely persisted as side chain fluctuations instead of backbone movements.[215] Site-specificity underlies all PTM effects on the ability for a protein to bind and function. In an enormous study on the effect of phosphorylation on G-protein-coupled receptor activation and arresting binding in 2020, the Dror group performed 1 ms of independent simulations to show that site-specificity of phosphorylation patterns results in different extents of activation and arrestin binding.[216] Including all other studies mentioned so far, each PTM position and type were shown to display unique effects per protein, which may not necessarily be confined to the neighborhood of the PTM site. And amazingly, decent extents of conformational change, and the conversion between different states, could be observed in very short amounts of simulation time.

What one should notice about this style of MD simulation and analysis is that its popularity as a design strategy has remained popular throughout the entire history of PTM simulations. That is, sufficient observations could be made from either classical or biased MD performed along shorter timescales. Conversely, these types of studies still have their limitations. Thermodynamic and kinetic insights are often not obtained from studies utilizing a "single" or "few" trajectory approach. However, the descriptive power of these studies is enabled because of the dramatic conformational changes caused by introduction of steric bulk or electrostatics by the specified PTM(s). Because of computational expense, a number of studies enumerated within this section simulated regions of proteins as peptides, which may affect how their findings translate to the effects PTMs have on their respective protein's global structure. In most of the exemplary literature cited in this section, a chief complaint written in their discussion sections was that not enough simulation could be performed because of available computing resources. Given how substantive conformational change could be seen even along the pico- to nano-second timescales, these foundational studies underscore the impact PTMs have on their local and distal environments. Still, one cannot help but wonder what could be seen from PTM simulations ranging from micro- to milli-second timescale trajectories, generating aggregate datasets, and/or implementing more advanced enhanced sampling methodology.

**5. Using aggregate datasets for the modulation of conformational free energy by posttranslational modifications**

Free energy landscape (FEL) representations of MD datasets are useful in that they represent a probability distribution as to how often a particular protein macrostate is observed given a finite extent of sampling. Because FELs represent distributions, they involve sampling to observe a given process. The extent of sampling required for a given process depends on its timescale, where rare events require enhanced sampling because they are higher energy. Rare events occur on slower timescales and are therefore less likely to record during unbiased simulation. Sufficient FEL coverage offers a comprehensive view of all possible conformations related to a protein's function (as described with a given order parameter set). Sampling rare events is then necessary to fully understand how a protein uses intermediate states as transitions between metastable states; note that these transitionary conformations are very difficult to resolve experimentally and can often only be observed computationally. Thus, time-independent data collection regimes using FELs of datasets gathered by enhanced sampling methods offer the unique opportunity to describe how PTMs affect entire protein conformational ensembles. The time frame for PTM studies implementing FEL-based approaches is primarily split across two time periods: the turn of the 21$^{st}$ Century and after 2015.

*5.1. Ensemble-based sampling of short timescale processes using classical molecular dynamics*
Depending on the timescale of the modeled process, PTM studies implementing a time-dependent research design can offer useful FEL-based insights. Examples of this include demonstrating how phosphorylation of a peptide shifts backbone torsion or dihedral angle conformational preferences.[196,197,199,217,218] Another process with a relatively short timescale includes dynamics of reasonably sized loops. Comparing 2.5 μs classical MD data of loop movement versus solvent accessible surface area of an E2 ubiquitin ligase enzyme showed that phosphorylated variants could only occupy an open, but not a closed, conformation because of the increased solvation required to satiate electronegativity along the installed phosphate groups.[219] FEL projection of principal component analyses for 1 μs WNK1 kinase simulations were also able to isolate specific activation loop conformations, demonstrating that a ~3 kcal/mol free energy barrier prevented the unphosphorylated activation loop from adopting an extended conformation readily seen in phosphorylated WNK1 simulations.[220] More stringent FEL analyses using dihedral angles further showed that unphosphorylated WNK1 data hardly had any conformational overlap with structures seen during phosphorylated simulations (Figure 7). However, the free energy difference between phosphorylated and unphosphorylated WNK1 in regions of shared dihedral angle conformational

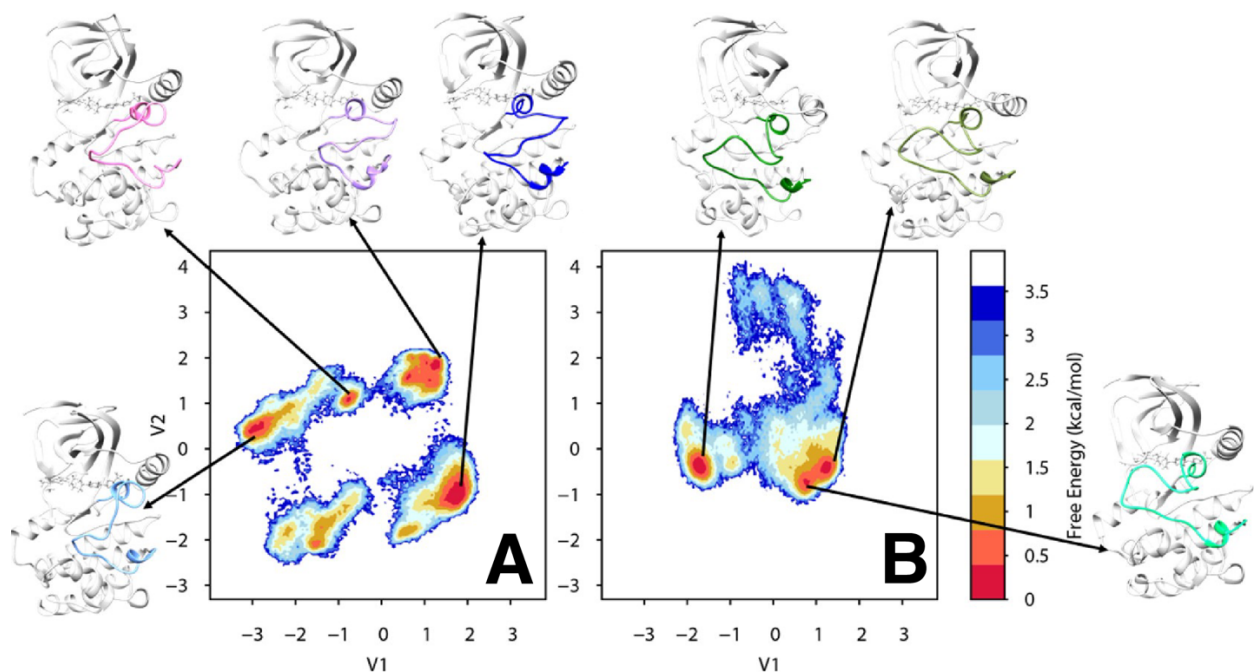

**Figure 7. Free energy landscapes for WNK1 kinase activation loop dynamics projected using order parameters derived from dihedral principal component analysis. (A)** Unphosphorylated WNK1. **(B)** Phosphorylated WNK1. Note that the minima for one proteoform's landscape is distinct from the other's, where any regions of overlapped sampling experience opposite favorability depending on phosphorylation state. Adapted with permission from N. A. Jonniya, M. F. Sk, and P. Kar, *ACS Omega*, 2019, **4**, 17404-17416.[220] Copyright 2022 American Chemical Society.

sampling still approximated ~3 kcal/mol.[220]

*5.2. Use of biasing potentials for simulating PTM dynamics*

When following single trajectory analyses, sometimes "vanilla" MD simulations performed at standard temperature and pressure conditions are insufficient for capturing processes associated with PTM-based conformational change. High temperature simulations have been used to facilitate unfolding of covalently modified proteins within the nanosecond timescale, suggesting how different helices stabilize binding interactions involving covalent modification by projecting helical unfolding FELs.[221] High temperature simulations can also facilitate FEL exploration by lowering transition barriers. Work from the Gervasio group done in 2017 combined high temperature unbiased MD with parallel tempering metadynamics to obtain more than 115 μs of total simulation data on p38α kinase activation upon phosphorylation.[222] Metadynamics simulations of up to 25 μs each for unphosphorylated, apo and holo dually phosphorylated p38α resolved conformational FELs were performed, where the order parameters used were distances distinguishing the active from inactive conformations (Figure 8). These p38α kinase FEL landscapes demonstrated that the unphosphorylated apo protein is unable to sample the active conformation, although phosphorylated apo protein can but at a much higher energetic cost. However, introduction of ATP-$Mg^{2+}$ encouraged deeper minima formation for phosphorylated p38α kinase surrounding an active conformation, whereas binding of ATP-$Mg^{2+}$ and the MK2 docking peptide drastically smoothened barriers between and deepened minima for phosphorylated intermediate and active states. That is, p38α kinase phosphorylation was shown to decrease the probability of inactive state conformations during metadynamics simulations. To confirm the thermo-stability or -instability of the metastable states seen for each system setup during metadynamics simulations, Gervasio and coworkers ran ten independent 1 µs-long classical MD simulations at 380 K from the respective p38α kinase crystal structures, extending simulation durations for any runs which happened to cross barriers into minima observed from metadynamics simulations. The independent unbiased runs showed strong agreement with the metadynamics results, as projecting the time-based progression of the independent data runs over the FELs showed that classical simulations resided in the aforementioned minima for longer periods of time.[221] While there was agreement between the biased and unbiased datasets, the conformational requirements for existing in a relatively active versus inactive conformation were further supported by FEL exploration from longer metadynamics trajectories. Otherwise, reliance on just the respective crystal structure conformations may have made comparisons between simulated states too stringent, possibly limiting how well phosphorylation was understood to impact extent of p38α activation.[221]

Other biased methods can be used to explore how PTMs affect transitions between metastable states. The impact of seven different combinatoric phosphorylation states on the activation and ATP-binding of plant receptor kinase BAK1 was explored via independent 500 ns Gaussian accelerated MD trajectories in 2020.[223] Comparing these phosphorylated proteoforms to unphosphorylated BAK1, the activation loop in unphosphorylated BAK1 was found to have the largest computed root-mean-square fluctuations. By contrast, the activation loop root-mean-square fluctuations for each of the other phosphorylated BAK1 proteoforms was reduced, suggesting a stabilizing role. Specifically, Thr455 phosphorylation appeared to be of secondary importance, further stabilizing the activation loop only when Thr450 was also phosphorylated. Comparison of each of the FELs, simulations involving pThr450 in any PTM combination always resulted in free energy minima surrounding an active BAK1 conformation. Thus, phosphorylation of Thr450 was suggested to be essential for stabilizing the BAK1 activation loop in an active conformation.

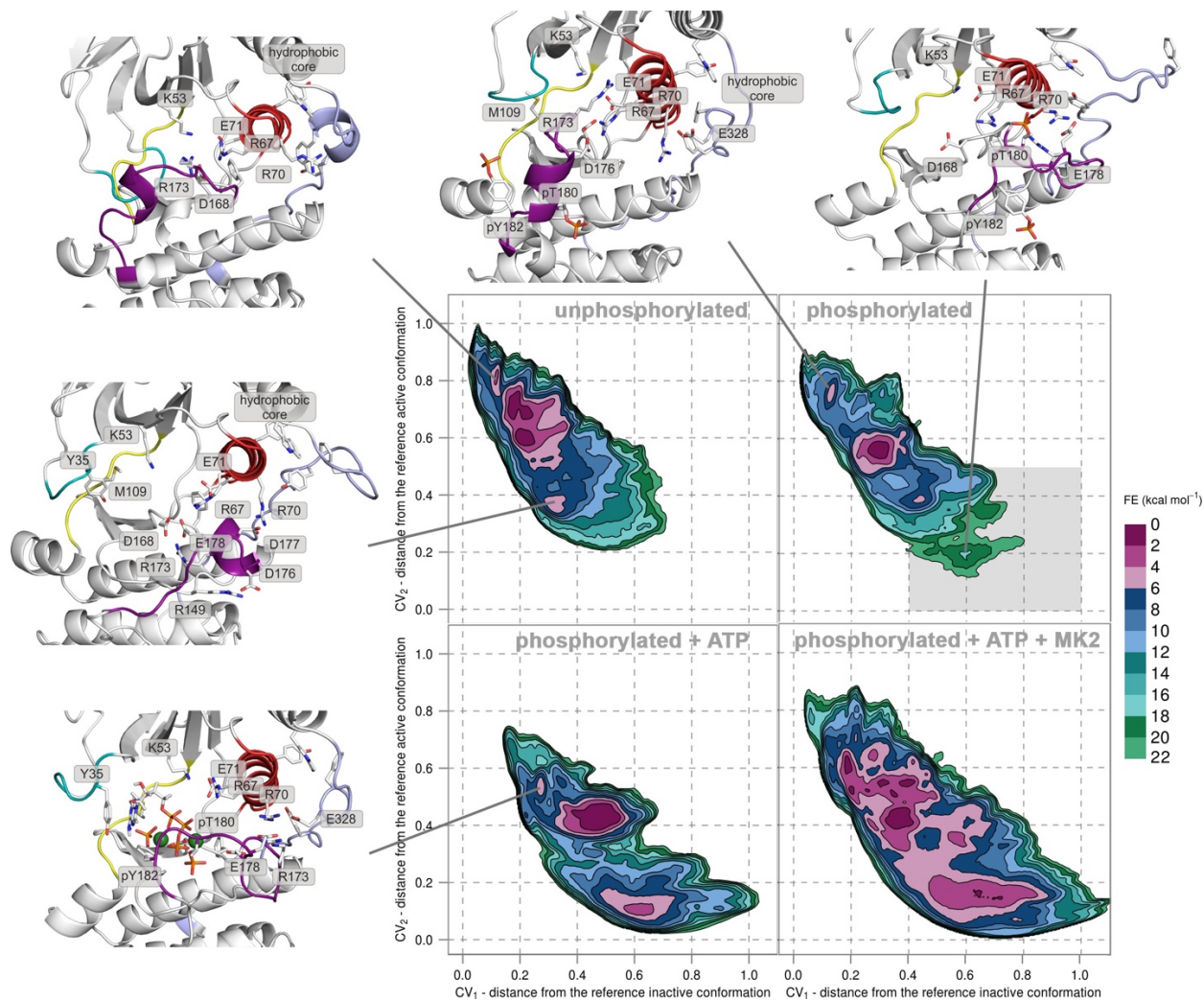

**Figure 8. Free energy surface for differentially phosphorylated p38α proteoforms.** The shared axes indicate functions representing the distance from the reference inactive (CV$_1$) and active (CV$_2$) structures for p38α. Representative structures from the reference publication are included to visualize differences between active- and inactive-like p38α states. Reproduced with permission from A. Kuzmanic, L. Sutto, G. Saladino, A. R. Nebreda, F. L. Gervasio and M. Orozco, *eLife*, 2017, **6**, e22175.[222] This article was originally published and distributed under the terms of the Creative Commons Attribution License, permitting unrestricted reproduction and adaptation provided proper crediting to author and source. Copyright 2022 eLife Sciences Publications Ltd.

Estimates of unbiased potentials of mean force through multistate Bennett acceptance ratio reweighting demonstrated that pThr450 uniquely engaged in salt bridges in each BAK1 phosphorylation scenario.[223]

Using the phosphorylated and unphosphorylated cystatin and NtrC crystal structures as case study proteins, the Wolynes group implemented an associated memory Hamiltonian for transferable structure prediction to model the effect of phosphorylation on each protein's folding FEL.[224] Here, the authors mutated the residue occupying the phosphorylation site into glutamic acid, where the Hamiltonian-based method enabled "supercharging" of the phosphorylation mimic. For cystatin, these Hamiltonian simulations predicted that phosphorylation makes transitions between the unfolded and folded states more energetically favorable for a simple two-state folding model (Figure 9A-B). Principal component analysis of cystatin snapshots from both the phosphorylated and unphosphorylated simulation datasets showed that transitionary semi-unfolded states were most similar, suggesting that cystatin is likely phosphorylated when occupying an intermediary structure somewhere closer to the unfolded state (Figure 8C). Meanwhile, NtrC FELs derived from the Hamiltonian simulations suggested that NtrC folding proceeds through a three-state folding model with a well-defined intermediate structure, regardless of phosphorylation status (Figure 6D-E). For NtrC, the largest extent of overlap for principal component analysis of simulation snapshots showed that the unfolded states for phosphorylated versus unphosphorylated states are most similar, suggesting that phosphorylation would likely occur on unfolded NtrC (Figure 8F).[224] Here, FELs enabled direct comparison between phosphorylated and unphosphorylated dynamics, where overlap in accessible conformations suggested which kinetic states could undergo phosphorylation *in cellulo*.

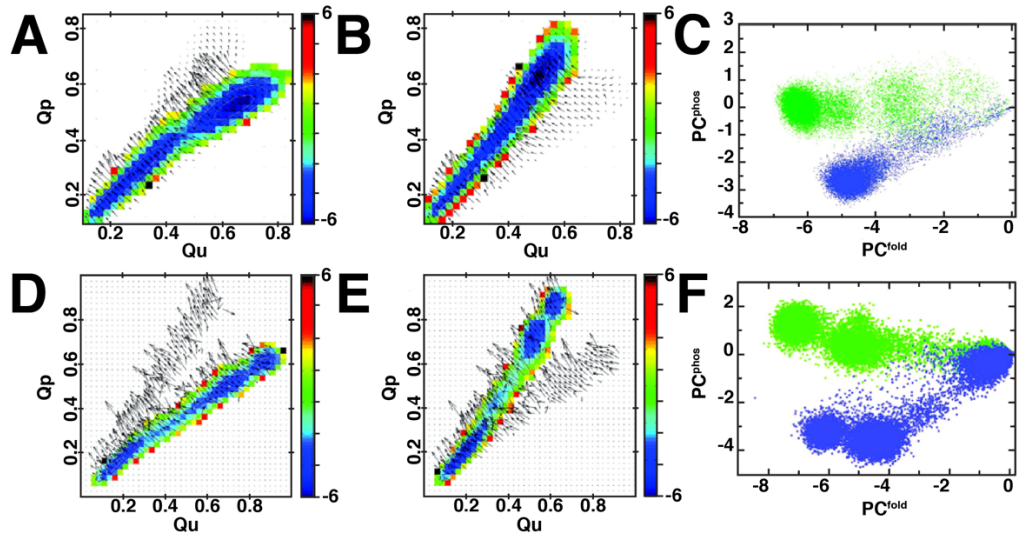

**Figure 9. Structure prediction Hamiltonian free energy landscapes of cystatin folding (A-C) and NrtC folding (D-F) in the phosphorylated and unphosphorylated forms.** Global order parameters reflecting how similar simulated structures are to the crystal structures for the phosphorylated (Qp) or unphosphorylated (Qu) proteoforms **(A-B, D-E)**. Principal component analysis (PCA) was performed to project the simulation data with respect to folded state ($PC^{fold}$) and phosphorylation state ($PC^{phos}$) **(C, F)**. For PCA plots, green data describes the unphosphorylated proteoform while purple data describes the phosphorylated proteoform. Cystatin folding was found to exhibit two-state behavior for the unphosphorylated **(A)** and phosphorylated **(B)** forms. Cystatin PCA suggested that phosphorylation could occur along a slightly unfolded transitionary state, due to the overlap in data sampled **(C)**. NrtC folding was found to exhibit three-state behavior for the unphosphorylated **(D)** and phosphorylated **(E)** forms. NrtC PCA suggested that phosphorylation must occur in a completely unfolded state based on overlap seen in the landscape **(F).** Adapted with permission from J. Latzer, T. Shen, and P. G. Wolynes, *Biochemistry*, 2008, **47**, 2110-2122.[224] Copyright 2022 American Chemical Society.

*5.3. Aggregate sampling for elucidating the impact of PTMs on conformational dynamics*

Another way to summarize a biomolecular process through FEL coverage is to run multiple parallel trajectories and then project the aggregate data against appropriate order parameters. Note that kinetic reweighting schemes (e.g., MSMs, milestoning) should accompany aggregate dataset projections, as the selection of seed structures for even classical MD trajectories does bias the resulting FEL.

*5.3.1. Ensemble-based sampling of short timescale processes using classical molecular dynamics*

Aggregate sampling approaches have been applied to study the effects of PTMs. One flavor of aggregate sampling studies is to run several trajectories from any given point and to then evaluate the total data based on some order parameters. For example, up to 5 μs of aggregate simulation for site-specific and additive villin headpiece carbonylation simulations were performed to examine the effects of carbonylation on villin thermostability in 2011.[225] Although time-dependent analyses demonstrated consistent increases in root-mean-square deviation; increased solvent accessible surface area within the hydrophobic core; and decreases in alpha-helical content as a result of carbonylation, the exact conformational effects differentiating villin headpiece native topology from carbonylated variations of villin could best be discerned using FEL projections.[225] Specifically, FELs comparing molecular hydrophobicity potential versus the sum of distances between phenylalanines occupying the villin hydrophobic core demonstrated that combinatoric carbonylations experienced a distinct metastable state, but could sample molecular hydrophobicity potentials and hydrophobic core compactness indicative of native or fully carbonylated minima. However, the opposite is not true, as the native and fully carbonylated villin structures predominantly occupied metastable states at either end of the distribution.[225]

*5.3.2. Markov state models on distinguishing PTM-based thermodynamic and kinetic changes*

Another flavor of aggregate sampling study designs is to use adaptive sampling to explore a conformational landscape. In adaptive sampling, the MD practitioner guides FEL exploration by running multiple, short, and parallel, unbiased trajectories in tandem as part of a "round" of simulation.[226–229] Seed structures for subsequent rounds are selected based on certain selection criteria that help facilitate efficient exploration of a landscape based on some order parameters.

Methodology has been proposed for making seed state selection as statistically unbiased as possible, as adaptive sampling can use any frame from a previous trajectory as a starting seed for the next round.[230–233] Because adaptive sampling regimes incorporate selection of multiple trajectories per round based on preexisting landscape coverage, they can improve statistical characterization of rare high energy regions between minima; traditionally, long trajectories struggle to capture rare events in more than one transition. Adaptive sampling workflows – and their representation through FELs – are easily amenable to Markov state model (MSM) generation.[40,177–179,234–240]

MSMs and adaptive sampling have been used to study the structural and regulatory effects of PTM on protein function. What matters most in these types of studies is that there is sufficient coverage of the FEL. In a 2017 study on the effect of S-glutathionylation on the plant receptor kinase BAK1, the Shukla group first used accelerated MD to accumulate ~22 μs of simulation data, from which starting structures were taken to seed an aggregate 132 μs of classical MD through an adaptive sampling workflow.[241] For this study, four simulation systems were constructed and compared: nonglutathionylated BAK1 core kinase domain (BAK1-SH) as a reference system, and then three singly glutathionylated on each possible glutathionylation site. After MSM construction, the authors quantified the global effects of S-glutathionylation on BAK1 conformational dynamics using the Kullback-Leibler divergence, a measure of similarity between two probability distributions.[242] These analyses suggested that one of the glutathionylation sites, Cys408, may serve as an inhibitory S-glutathionylation site responsible for the decrease of activity in ABK1. Firstly, glutathionylation on C408 caused dramatic effects on BAK1 structure throughout the kinase domain, whereas little changes were observed during the other glutathionylation simulation datasets. Furthermore, C408 glutathionyation shifted the free energy landscapes away from the active-like state, while modifications of other sites retained most features of the unmodified protein. Here, adaptive sampling-based exploration of the BAK1 kinase activation FEL provided sufficient input data for explaining how S-glutathionylation allosterically controlled kinase activity.[241]

The Shukla group constructed an MSM from ~320 μs of aggregate MD data on plant abscisic acid (ABA) receptors to delineate the role of tyrosine nitration ABA signaling inhibition during 2019.[243] Comparing the conformational FELs of the PYL5 receptor with and without nitration revealed that ABA cannot bind to PYL5 after tyrosine nitration (Figure 10). Instead, ABA was stabilized by multiple hydrogen bonds formed by residues surrounding one of the nitrotyrosine residues. Tyrosine nitration was found to significantly shift nearby residue backbone positions when compared to unmodified PYL5. Collectively, tyrosine nitration of the PYL5 receptor rearranged the binding pocket to prevent ABA from reaching the final binding site and therefore inhibiting ABA signaling.[243] Although MSMs offer a kinetic framework based on the aggregate stitching of individual trajectories, their resulting transition probabilities calculated for interconversion between different clustered microstates can be used to instantiate a single trajectory. Often sampled using Monte Carlo simulations along the MSM, a single unbiased trajectory can then be generated which directs a biomolecular system from one conformation to another, all while simultaneously capturing each of the transitions identified along the specified path. The Shukla group's PYL5 receptor study used Monte Carlo sampling to generate a 3 ms trajectory depicting unmodified PYL5 activation.[243] While the same analysis was not performed on nitrated PYL5 because its FEL did not sample an activated state, trajectories formed from MSM-based Monte Carlo sampling could be used to consolidate data from aggregate sampling for time-dependent analyses.

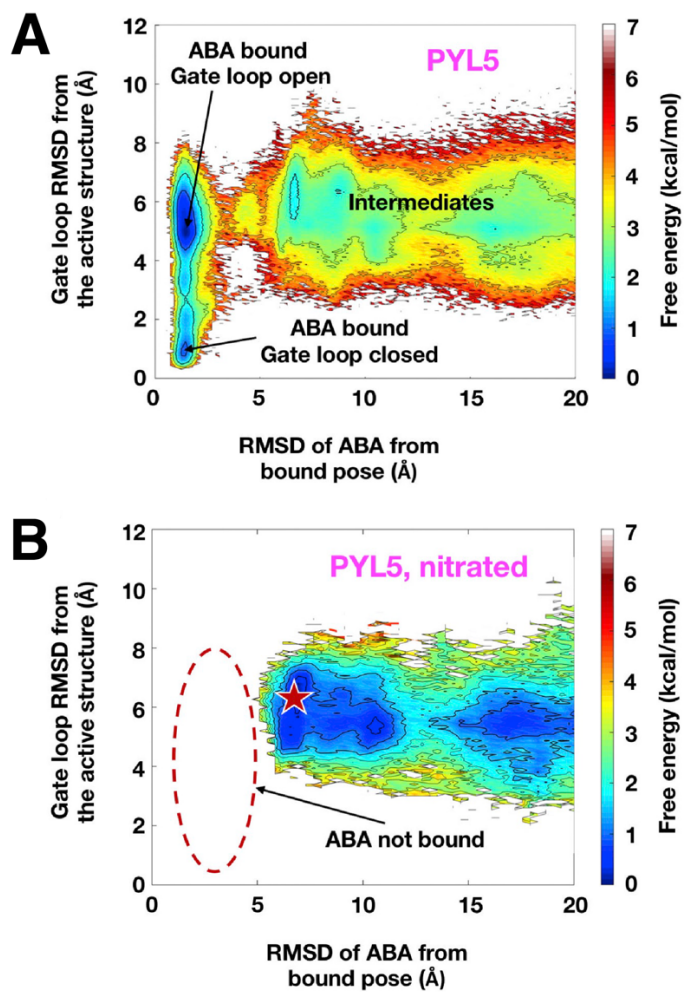

**Figure 10. The effects of tyrosine nitration on ABA binding.** The free energy landscapes of ABA binding to the **(A)** unmodified and **(B)** nitrated PYL5 receptors. The structure closest to the active ABA-bound pose, labeled as a red star in **(B)**. Adapted with permission from S. Shukla, C. Zhao, and D. Shukla, *Structure*, 2019, **27**, 692-702.e3.[243] Copyright 2022 Elsevier.

In addition to demonstrating how PTMs perturb the likelihood of a certain conformation being accessed, MSMs can also provide information about the timescale for kinetic transition between different macrostates observed along a FEL. The Shukla group did so by implementing transition path theory in their NRT1.1 phosphorylation study.[244,245] The plant nitrate transporter NRT1.1 is regulated by phosphorylation. With affinity inverse to the levels of environmental nitrate, NRT1.1 is either a low-affinity transporter during saturating nitrate conditions or a high-affinity transporter during desaturating nitrate conditions.[246] Additionally, the phosphorylation state determines the oligomeric state of NRT1.1, as the transporter is known to dimerize when in the unphosphorylated state.[247,248] To explore the effects of phosphorylation on enhanced transport and transporter oligomerization, the Shukla group performed unbiased MD simulations for four systems: unphosphorylated NRT1.1 (UnpNRT1.1) (~9 μs) and phosphorylated (pNRT1.1) (~5 μs) dimer, as well as UnpNRT1.1 (~142 μs) and pNRT1.1 (~63 μs) monomer.[244] Cross-correlation analyses among all Cα atoms within each system demonstrated that dynamic coupling disappeared at the dimeric interface for pNRT1.1 versus UnpNRT1.1 dimer (Figure 11A-C). Meanwhile, dynamic correlation within each monomer in pNRT1.1 was enhanced compared with the UnpNRT1.1 dimer. Thus, phosphorylation appeared to decouple the dimer and allow the two monomers to behave independently. Increasing independent behavior by each participating monomer enhanced their respective structural flexibility, leading to a higher transport rate. Using transition path theory calculations, the average time required for one complete cycle from inward-facing to outward-facing conformations was 5 ± 2 μs and 18 ± 3 μ for pNRT1.1 and UnpNRT1.1 monomer, respectively. FELs further reflect these differences in transport rate (Figure 11D-E). UnpNRT1.1 and pNRT1.1 monomer FELs show how phosphorylation stabilized the outward-facing state while eliminating unnecessary intermediate state, thereby facilitating transport by hastening the transition from inward-facing to outward-facing states. Overall, this study suggested that phosphorylation accelerates necessary conformational transitions, resulting in a higher transport rate.[244]

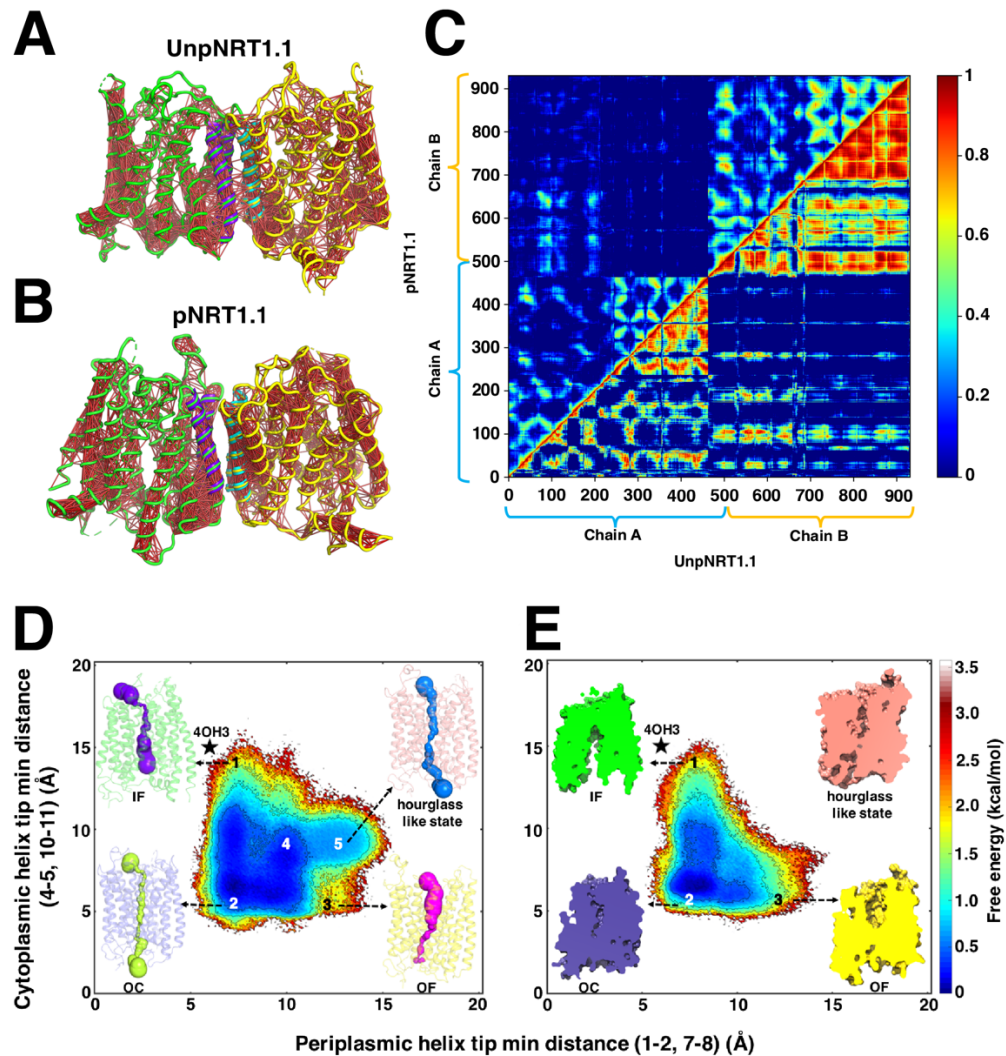

**Figure 1. The effects of phosphorylation on NRT1.1. (A,C)** Dynamic cross-correlation matrix (DCCM) analysis for all alpha carbons in UnpNRT1.1 and **(B,C)** pNRT1.1 dimer. DCCM results displayed in loop representation for **(A)** and **(B)**. Dimer interface is colored purple and cyan for Chain A and B, while the remaining parts are colored in green and yellow, respectively. Red lines linking residue pairs represent strong dynamic pairwise coupling between residues. **(D)** Conformational free energy landscapes for UnpNRT1.1 monomer. **(E)** Conformational free energy landscapes for pNRT1.1 monomer. Intermediate states are labeled as (1) inward-facing (IF), (2) occluded (OC), (3) outward-facing (OF), (4) partial IF-OF (inward-outward facing), and (5) hourglass-like state, respectively. Adapted with permission from B. Selvam, J. Feng, and D. Shukla, *bioRxiv*, 2022.[244] Permission directly granted from authors for use of preprint figures.

*5.4. Summary of PTM simulations using ensemble-based approaches for free energy landscape construction*

Studies implementing aggregate sampling approaches to cover a FEL still fundamentally provide similar insights as studies strictly following single trajectories and their time-dependent analyses. As the Shukla group constructed MSMs on over 600 μs of aggregate simulation data to demonstrate that Thr276-phosphorylated human serotonin transporter experienced lower relative free energy barriers for transitions between occluded to inward-facing states, the basis for this lower free energy barrier was due to rearrangements in nearby hydrogen-bond networks lining the intracellular gate.[249] Yes, time-dependent approaches to studying PTM with MD simulation have declared similar results; changes in hydrogen bonding networks or sidechain orientations are indeed the immediate response mechanism to local PTM installation, where these changes may allosterically communicate the need for long-range structural rearrangements as well. But what separates previous studies from those implementing aggregate sampling and/or MSMs is that the latter works are able to define what constitutes a specific conformational state, as well as identify the path by which one conformational state transitions into another. That is, aggregate simulation studies can evaluate the thermostability of different conformations through more rigorous statistical sampling. Based on the ergodic assumption that all MD trajectories will eventually sample the same phase space for a given biomolecular process, aggregate sampling just uncovers a greater extent of the stationary density for the process under inspection. MSMs, however, also provide kinetic insights on the evolution of a given molecular process. A specific PTM may not alter the phase space but instead change the kinetics required to traverse across FELs. Trajectory-driven results can offer a relative perspective on the residence time for a given molecular event, but are otherwise unable to provide kinetic rates for transitioning between metastable states separated by larger extents of conformational change. Said calculations do come at a cost, requiring more involved simulation and analyses workflows, often accompanied with greater computational resource needs. Additionally, uncertainty estimates and other postprocessing for statistical validation are necessary to ensure convergence around ensemble averages.[41,48]

## 6. Outlooks on PTM modeling and simulations

It has proven difficult to say exactly how any given PTM universally affects some proteins' structures and functions. Experimental throughput makes characterization at atomistic resolution difficult, leaving much of the burden to molecular modeling for explaining cause-and-effect relationships between site-specific PTMs, structural changes, and altered protein function. Luckily, molecular modeling and simulation are perfectly poised to embrace the challenge of PTM characterization laid ahead. Structural and informatic modeling of PTMs in kinases has shown their phosphorylation sites to be relatively conserved, suggesting a potential for *in silico* PTM characterizations to at least universally apply to proteins within the same family.[250] Still, proteins and their covalently modified proteoforms need to be systematically studied so that the relationship between regulatory roles and structural effects of PTMs can be fundamentally understood. Much of the advances seen in machine learning capabilities have been applied towards the prediction and identification of PTMs and their respective sites for covalent modification. As the input data for PTM prediction and identification is based on mass spectrometry results, machine learning-based predictions on functional characterization will require larger datasets focused on characterizing PTM function. In this sense, molecular dynamics simulations are unrivaled in their ability to generate large swaths of data from which distributions of molecular descriptors can be compared to covalent modification status using regression-based models.

Changing focus from simulation outputs, innovation in computing capabilities means that simulation inputs and their parameterization can also be expected to improve. PTM chemical diversity must be appropriately covered by all force field sets for more amenable simulated applications to any research problem. The GROMOS force field library offers tremendous coverage of PTMs. Still, MD practitioners gravitate towards specific force field libraries when tackling specific research problems. For instance, plans to simulate membrane proteins may result in CHARMM force field parameterization given its fantastic coverage of lipid diversity with over 2000 lipid molecules in different starting conformations for lipid bilayer building.[251,252] Still, CHARMM parameterization with compatible force fields for less common PTMs are not necessarily automatically, nor easily, incorporated into existing computational tools as is seen per the GROMOS library.[9,253] As observed in the development of a membrane protein-specific OPLS-AA force field,[254] it is only a matter of time before force field libraries recognize their application-specific deficiencies and offer the same research opportunities for modeling PTMs with equivalent ease of access. That is, until efforts like the Open Force Field Initiative and machine learning-based force field libraries can consistently offer the accuracy of *ab initio* methods without incurring the same computational cost.[255–257] The same argument applies to the efficient use of polarizable force fields when dealing with PTM protonation state.[258,259]

Aside from parameterization improvements, advances in structure prediction methodology raise the possibility of predicting covalently modified proteoform conformations. Software like DeepMind's AlphaFold, along with the Baker lab's RoseTTaFold, have now revolutionized protein structure prediction with their machine learning-based template-free methods that, in most cases, can rival experimental structure quality.[260–263] How these software are learning from numerous multiple sequence alignments and pairwise evolutionary constraints to generate a distribution of probable models suggests that incorporating the effects of PTMs in state-of-the-art conformer prediction is not so farfetched. Previous efforts using molecular mechanics showed that it was possible to predict protein conformational change upon covalent modification given an unmodified structure.[264] Now, tools like Privateer have already been developed to graft glycan PTMs directly onto AlphaFold predicted structures.[265] Inter-residue impacts of PTM-containing structures in the Protein Data Bank could be learned and included as a stage within these state-of-the-art structure prediction pipelines.[266] Given how a majority of PTMs (excluding ubiquination) occur along intrinsically disordered regions of proteins, integrated modeling of PTMs onto even difficult-to-predict protein structures is more feasible than ever because of AlphaFold.[267] Including post-predictional modifications could offer a unique opportunity for seeding MD simulations from more unique starting structures for more efficient PTM characterization.

Regardless of the inputs provided, future PTM simulation studies that will capture vast free energy landscape coverage should strive to standardize analyses as to leverage their extensive sampling of macrostate population densities. As PTM installment can mirror the effects of mutation in shifting free energy landscapes, performing routine allosteric analyses on vast PTM simulation datasets could make the interpretation of long-range effects of PTMs more systematic across different proteins.[268–271] PTM studies incorporating MSMs should report the exact transition probability matrix weights and calculate macrostate kinetic flux rates via transition path theory.[177–179,245] In doing so, how often a PTM proteoform exists in a specific conformation (e.g., active versus inactive) can be ascertained. Offering probability distributions relative to the specific activity of a protein after PTM installation could drastically improve the accuracy of systems-scale kinetic analyses for multiscale modeling endeavors.[1,272–274] State-of-the-art structure predictions and faster computing hardware now unleash an array of *in silico* opportunities to realistically characterize PTMs for proteins that lack experimentally resolved structures or are simply understudied.[275] Together, modern MD simulations can join theory and experiments in informing us as to how the structure-dynamic perturbations caused by PTMs impact whole-cell physiology.[273,274]

Despite the opportunities they provide in characterizing PTMs, it is worth recognizing the caveats of molecular simulations. A major caveat to molecular simulations is that the process of covalent bond breakage and formation is extremely expensive to model. This reality is a major driver for the development of machine learning-based force fields so that molecular mechanics simulations can be performed with parameterization done at the accuracy level of quantum mechanics simulations. Thus, simulations are likely unable to provide as valuable insights as experiments when it comes to dynamical modeling for the direct instance of PTM, such as the rate of covalent modification or cellular trafficking. Experimental characterization can further delineate spatio-temporal factors underlying PTM effects which are not easily, if entirely possible to be, modeled using MD simulations. What is a simple task for computational research is often an arduous one for experiments, and vice versa. Still, the level of control and atomistic resolution offers a glimpse as to what could be happening given explicit conditions and sufficient sampling.

Throughout this perspective, we have highlighted almost 30 years' worth of PTM research using *in silico* molecular modeling and simulation techniques. Much of this work has demonstrated how molecular dynamics simulations can characterize the structural consequences induced by PTMs.

With recent computing developments and simulation methodologies enabling MD to be more efficient than ever before, the future of PTM-based molecular dynamics research can more easily employ currently established enhanced sampling and structure prediction methodology. Computational research efforts surrounding PTMs and their modeling will shift towards identifying new conformations caused by PTMs, and the frequency at which these conformations are sampled within the stationary density for the specified proteoform. Systematic computational characterization leveraging larger extents of sampling will lead to systematic translational gains in experimental efforts surrounding therapeutic developments and bioengineering. The predictive power of MD simulations is limited insofar as the realism used to model the system. As efforts have been cast to make the modeling of lipid bilayers and membrane proteins match experiments with improved realism,[276] so too shall efforts surrounding PTM simulation strive to make our biophysical understanding of processes regulating protein function more complete.

## Acknowledgments

The authors would like to thank Matthew Chan for providing comments on earlier drafts of this manuscript. Other Shukla Group members – Prateek Bansal, Jiming Chen, Tanner Dean, Soumajit Dutta – are thanked for recommending additional references for the completion of this manuscript. D.S. acknowledges support from the Foundation for Food and Agriculture Research via the New Innovator Award and NSF Early CAREER Award (NSF-MCB-1845606), as well as the National Institutes of Health (Award No. R35GM142745).

## Conflicts of Interest

The authors declare no conflicts of interest.